\documentclass[10pt,preprintnumbers, floatfix, letterpaper, onecolumn,aps,prd,epsfig,nofootinbib,nat_bib,longbibliography]{revtex4-1}

\usepackage{graphicx}
\usepackage[utf8]{inputenc}
\usepackage[T1]{fontenc}
\usepackage{physics}
\usepackage{latexsym}
\usepackage{epstopdf}
\usepackage{latexsym}
\usepackage{amssymb}
\usepackage{amsmath}
\usepackage{mathtools}
\usepackage{color}
\usepackage{mathrsfs}
\usepackage{xparse}
\usepackage{enumerate}
\usepackage{multirow,tabularx,boldline,array}
\usepackage{bbding}
\usepackage{enumitem}
\usepackage{pifont}
\usepackage[center]{subfigure}
\usepackage[
            pdfstartview=FitH,
            bookmarksnumbered=true,
            bookmarksopen=true,
            colorlinks,
            linkcolor=blue,
            anchorcolor=green,
            citecolor=blue
            ]{hyperref}
\begin{document}

  \renewcommand\arraystretch{2}
 \newcommand{\bq}{\begin{equation}}
 \newcommand{\eq}{\end{equation}}
 \newcommand{\bqn}{\begin{eqnarray}}
 \newcommand{\eqn}{\end{eqnarray}}
 \newcommand{\nb}{\nonumber}
 \newcommand{\lb}{\label}
 \newcommand{\cb}{\color{blue}}
    \newcommand{\cc}{\color{cyan}}
        \newcommand{\cm}{\color{magent_a}}
\newcommand{\rc}{\rho^{\scriptscriptstyle{\mathrm{I}}}_c}
\newcommand{\rd}{\rho^{\scriptscriptstyle{\mathrm{II}}}_c}
\NewDocumentCommand{\evalat}{sO{\big}mm}{%
  \IfBooleanTF{#1}
   {\mleft. #3 \mright|_{#4}}
   {#3#2|_{#4}}%
}
\newcommand{\PRL}{Phys. Rev. Lett.}
\newcommand{\PL}{Phys. Lett.}
\newcommand{\PR}{Phys. Rev.}
\newcommand{\CQG}{Class. Quantum Grav.}


\title{A Note on Entanglement Entropy for Primary Fermion Fields in JT Gravity}
\author{Chang-Zhong Guo}
	\email{chang.zhong.guo1997@gmail.com}
	\affiliation{Department of Physics, Nanchang University, Nanchang, 330031, China}
	\author{Wen-Cong Gan}
	\email{Wen-cong$\_$Gan1@baylor.edu}
	\affiliation{GCAP-CASPER, Physics Department,
Baylor University, Waco, Texas 76798-7316, USA}
	\affiliation{Department of Physics, Nanchang University, Nanchang, 330031, China}
		\author{Fu-Wen Shu}
	\email{shufuwen@ncu.edu.cn; Corresponding author}
	\affiliation{Department of Physics, Nanchang University, Nanchang, 330031, China}
	\affiliation{Center for Relativistic Astrophysics and High Energy Physics, Nanchang University, Nanchang,
330031, China}
\affiliation{GCAP-CASPER, Physics Department,
Baylor University, Waco, Texas 76798-7316, USA}
\affiliation{Center for Gravitation and Cosmology, Yangzhou University, Yangzhou, China}

\date{\today}

\begin{abstract}

In this paper we analyse and discuss 2D Jackiw-Teitelboim (JT) gravity coupled to primary fermion fields in asymptotically anti-de Sitter (AdS) spacetimes. We get a particular solution of the massless Dirac field outside the extremal black hole horizon and find the solution for the dilaton in JT gravity. Two dimensional JT gravity spacetime is conformally flat, we calculate the two point correlators of primary fermion fields under the Weyl transformations. The key point of this work is to present a standard technique which is called {\it resolvent} rather than CFT methods. We redefine the fields in terms of the conformal factor as the fermion fields, and we use the resolvent technique to derive the renormalized entanglement entropy for  massless Dirac fields in JT gravity.

\end{abstract}

\maketitle
\tableofcontents
\section{Introduction}
 \renewcommand{\theequation}{1.\arabic{equation}}\setcounter{equation}{0}

  Two dimensional JT gravity \cite{Teitelboim:1983ux,Jackiw:1984je,Jackiw:1991nb} is a model of 2D dilaton gravity which admits AdS$_{2}$ holography \cite{Almheiri:2014cka}, also it is the simplest nontrivial theory of gravity. In recent years, JT gravity has provided a simple and meaningful toy model for the study of black hole information loss problem. In particular, it has been able to describe the Page curve of black hole entropy, which is a key step towards solving black hole information paradox \cite{Almheiri:2019psf,Almheiri:2019yqk,Hollowood:2020cou}. All these works suggest that after the Page time, there is a configuration that the entanglement wedge of Hawking radiation include an island inside the black hole interior, and the island configuration is the key to reproducing the Page curve. Therefore, it is of great significance to verify the validity of the island configuration. Motivated by this, there have been several proposals recently to show the existence of the island by proposing ways to extract information from the island to the radiation \cite{Faulkner:2018faa,Cotler:2017erl,Chen:2019gbt,Penington:2019kki}. One of them is achieved by making use of the modular Hamiltonian and modular flow in entanglement wedge reconstruction and the equivalence between the boundary and bulk modular flow \cite{Chen:2019iro}. As a concrete example, they consider extremal black holes with modular flow in JT gravity coupled to baths. They claim that the explicit information extraction process can be observed in the case that the bulk conformal fields contain free massless fermion fields \cite{Chen:2019iro}.

While the proposal in \cite{Chen:2019iro} shows a promising way to extract information from the island configuration in JT gravity, the details of this process have not been fully specified in the literature. In particular, the modular flow of the free massless fermion field considered in \cite{Chen:2019iro} is in two dimensional Minkowski spacetime. More details are needed on how to apply this flow to the conformally flat spacetime. Therefore, in this paper, we aim to fill this gap in the literature by providing detailed calculations of entanglement entropy for massless fermion fields with the help of the resolvent technique. Our goal is to provide a clear and comprehensive understanding of the proposed method and its implications for the black hole information paradox.

%
%

  This paper is organized as follows. In Sec.\ref{sec-entropy}, we get the equations of motions in the background of JT gravity coupled to primary fermion fields and we find the particular solution of the wave function outside the extremal black hole horizon, and we also solve for the dilaton in JT gravity. In Sec.\ref{4d-far}, we calculate the two point correlators of primary fermion fields under Weyl transformations by CFT method. In Sec.\ref{4d-near}, we review a standard technique called {\it resolvent} to derive the entanglement entropy in $n$ disjoint intervals for a massless Dirac field in two dimensional vacuum Minkowski spacetime \cite{Casini:2009vk,Casini:2009sr}. Correspondingly, we redefine the fields in terms of the conformal factor as the fermion fields, and we use the resolvent technique as described in two dimensional vacuum Minkowski spacetime to derive the renormalized entanglement entropy for  massless Dirac fields in JT gravity.

\section{Primary fermion fields in JT gravity background }\lb{sec-entropy}
 \renewcommand{\theequation}{2.\arabic{equation}}\setcounter{equation}{0}

The JT gravity model consists of 2D gravity coupled to a scalar $\phi$ called the dilaton, with a classical bulk term action in Lorentzian signature on an asymptotically AdS spacetime,
\bqn\lb{bjt}
S_{\text{JT}}=\frac{1}{16\pi G_N}\int d^{2}x\sqrt{-g}\left(\phi R+2\phi-2\phi_0\right),
\eqn
where $R$ is the Ricci scalar and we have set the AdS$_{2}$ length $l_{AdS} = 1$. The JT gravity action originates from a dimensional reduction
of four dimensional near extremal magnetic charged black hole \cite{Nayak:2018qej,Fabbri:2005mw,Grumiller:2002nm}, the two-dimensional JT model is obtained by reduction of the spherically symmetric metric,
\bqn
ds^{2}_4=g_{\mu\nu}(t,r)dx^{\mu}dx^{\nu}+\phi(t,r)d\Omega^{2}_2,
\eqn
where $g_{\mu\nu}$ is the 2D part with coordinates $(t,r)$ and the dilaton $\phi$ plays the role of the radius of the 2-sphere which we want to reduce, and $\phi_0$ is a constant which is proportional to the extremal entropy of the higher-dimensional black hole geometry.

In this paper, we consider the coupling of a massless Dirac field $\Psi(x)$ to JT gravity. The massless Dirac field is also called the primary field which satisfies conformal invariance under conformal transformations in CFT method. The action of primary fermion fields in 2D curved spacetime is \cite{Freedman:2012zz,Collas:2018jfx,Lippoldt:2016ayw,Guendelman:2012sq,Parker:2009uva}:

\bqn\lb{Dirac}
S_{\text{D}}=\frac{i}{2}\int d^{2}x\sqrt{-g} \overline{\Psi}\left(\overline{\gamma}^{\mu}\overleftrightarrow{D_{\mu}}
\Psi\right),
\eqn
where $\overrightarrow{D_{\mu}}=\overrightarrow{\partial_{\mu}}+\Gamma_{\mu}=\overrightarrow{\partial_{\mu}}+\frac{1}{8}\eta_{ac}{{\omega_{\mu}}^{c}}_{b}[\gamma^{a},\gamma^{b}]$ is the spinor covariant derivative, and the spin
connection is ${{\omega_{\mu}}^{c}}_{b}= -{e_{b}}^{\nu}\left(\partial_{\mu}{e^{c}}_{\nu}-\Gamma_{\mu\nu}^{\lambda} {e^{c}}_{\lambda}\right)$\footnote{A tetrad is a set of linearly independent vectors that can be defined at each point in a Riemannian spacetime,
the tetrads by definition satisfy the relations:
${e^{a}}_{\mu} {e_{a}}^{\nu}=\delta_{\mu}^{\nu},\ {e^{a}}_{\mu} {e_{b}}^{\mu}=\delta^{a}_{b}$. The choice of the tetrad field determines the metric:
$g_{\mu \nu} ={e^{a}}_{\mu} {e^{b}}_{\nu} \eta_{ab}, \ \eta_{ab} ={e_{a}}^{\mu} {e_{b}}^{\nu} g_{\mu \nu}$.}.
Note that in eq.(\ref{Dirac}), $i\overline{\Psi}\left(\overline{\gamma}^{\mu}\overrightarrow{D_{\mu}}\Psi\right)$ is not real, so we should choose $\frac{i}{2}\overline{\Psi}\left(\overline{\gamma}^{\mu}\overleftrightarrow{D_{\mu}}\Psi\right)$ as the Dirac Lagrangian, where $\overleftarrow{D_{\mu}}=\overleftarrow{\partial_{\mu}}-\Gamma_{\mu}=\overleftarrow{\partial_{\mu}}-\frac{1}{8}\eta_{ac}{{\omega_{\mu}}^{c}}_{b}[\gamma^{a},\gamma^{b}]$ operates on $\overline{\Psi}$, and $\overleftarrow{D_{\mu}}$ is different from $\overrightarrow{D_{\mu}}$.

We adopt the metric signature $(-,+)$ and the anticommutator of the Dirac gamma metric is $\{\gamma^{a},\gamma^{b}\}=2\eta^{ab}\textbf{\textit{1}}$. The Dirac gamma matrices have this property: $(\gamma^{0})^{2}=-\textbf{\textit{1}}$ and $(\gamma^{1})^{2}=\textbf{\textit{1}}$, we choose
\bqn
\gamma^{0}=\left(\begin{array}{cc}0 & 1\\-1 & 0\end{array}\right), \quad\gamma^{1}=\left(\begin{array}{cc}0 & 1 \\1 & 0\end{array}\right).
\eqn
The Dirac adjoint in eq.(\ref{Dirac}) is defined as $\overline{\Psi}=\Psi^{\dagger}\gamma^{0}$ , and $\overline{\gamma}^{\mu}={e_{a}}^{\mu}\gamma^{a}$, where ${e_{a}}^{\mu}$ is the vierbein.

We define $\alpha$ as the strength of the coupling between the massless Dirac field and JT gravity, and we also define $\kappa^{2}\equiv 8\pi G_N$, then the total action functional is
\bqn\lb{st}
S=S_{\text{JT}}+\alpha S_{\text{D}}=\int d^{2}x\sqrt{-g}\Big[\frac{1}{2\kappa^{2}}\left(\phi R+2\phi-2\phi_0\right)+\frac{i\alpha}{2}\overline{\Psi}\left(\overline{\gamma}^{\mu}\overleftrightarrow{D_{\mu}}\Psi\right)\Big].
\eqn
By varying the total action \eqref{st} with respect to the metric field, then we get the classical equations of motion (see Appendix (\ref{app-1})):
\bqn\lb{Equation}
g_{\mu \nu}(\phi-\phi_0)+\nabla_{\mu}\nabla_{\nu}\phi-g_{\mu \nu}\square\phi=\frac{i\alpha\kappa^{2}}{8}\overline{\Psi}\Big(\gamma_{\nu}\overleftrightarrow{D_{\mu}}+\gamma_{\mu}\overleftrightarrow{D_{\nu}}\Big)\Psi,
\eqn
where $\gamma_{\nu}$ is defined as $\gamma_{\nu}=(g_{\mu \nu}{e_{a}}^{\mu})\gamma^{a}=g_{\mu \nu}\overline{\gamma}^{\mu}$, and $i\overline{\Psi}\left(\gamma_{\mu}\overleftrightarrow{D_{\nu}}\Psi\right)$ is defined as $i\overline{\Psi}\left(\gamma_{\mu}\overleftrightarrow{D_{\nu}}\Psi\right)=i\overline{\Psi}\gamma_{\mu}\overrightarrow{D_{\nu}}\Psi+
\left(i\overline{\Psi}\gamma_{\mu}\overrightarrow{D_{\nu}}\Psi\right)^{\dagger}$, with $\left(i\overline{\Psi}\gamma_{\mu}\overrightarrow{D_{\nu}}\Psi\right)^{\dagger}=-i\left(\overline{\Psi}\overleftarrow{D_{\nu}}\right)\gamma_{\mu}\Psi$.
\subsection{Massless Dirac fields outside the extremal black hole horizon}
\begin{figure}[h]
\includegraphics[height=7.5cm]{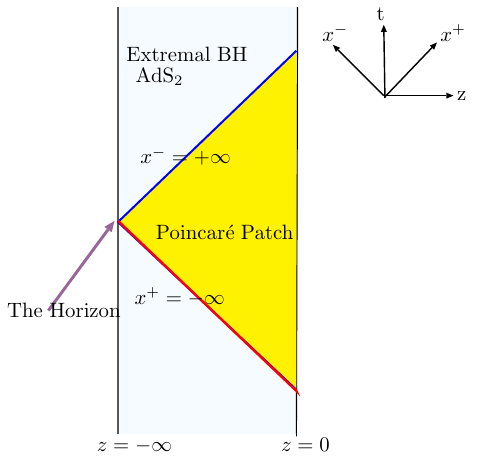}
\caption{The Penrose diagram for the extreme black hole in JT gravity. The yellow region is the Poincar\'{e} patch where the wave function is distributed in. The blue null line is the future event horizon and the red null line is the past event horizon. Here $z$ ranges from $z \in (-\infty,0]$, where $z = -\infty$ is the location of the horizon.}
\label{5}
\end{figure}

In a generic conformal coordinate system $x^{\pm}$, the metric in two dimensional gravity is given by
\bqn\lb{388}
ds^2=-e^{2\rho(x^+,x^-)}dx^+ dx^-.
\eqn
In this paper we consider a zero temperature black hole in the two-dimensional Jackiw-Teitelboim gravity, and we can use the Poincar\'{e} coordinates $x^{\pm} = t \pm z$ to describe the extremal black hole (see the Fig.\ref{5} for more details). The metric in the Poincar\'{e} patch is
\begin{eqnarray}
ds^{2}=-\frac{4dx^+ dx^-}{\left(x^+-x^-\right)^2}=\frac{-dt^{2}+dz^{2}}{z^{2}} , \; z \leqslant 0.
\end{eqnarray}
 The boundary of AdS$_{2}$ spacetime is at $z = 0$, the future horizon of the JT extremal black hole is at $x^{-}=+\infty$, while the past horizon is at $x^{+}=-\infty$.

 By varying the Dirac action $S_{\text{D}}$ with respect to the Dirac field, we can get the massless Dirac field equation in two dimensional conformally flat spacetime
 \bqn
 i\overline{\gamma}^{\mu}{D_{\mu}}\Psi=0.
 \eqn
 We can write the 2-component massless Dirac spinor $\Psi$ as
 \bqn\lb{gcz}
 \Psi=\left(\begin{array}{cc}\Psi_{1} \\ \Psi_{2}\end{array}\right)=\left(\begin{array}{cc}\psi_{1}+i\psi_{2} \\ \psi_{3}+i\psi_{4}\end{array}\right).
 \eqn

 Any two dimensional spacetime is conformally flat, the massless Dirac field equation in the conformal gauge can  be written as\footnote{A tetrad is a set of four linearly independent vectors that the direction can be arbitrarily selected, four vierbeins are constrained by three equations in light cone coordinates: $\eta_{00}=-1=2g_{+-}{e_{0}}^{+}{e_{0}}^{-}$, $\eta_{11}=1=2g_{+-}{e_{1}}^{+}{e_{1}}^{-}$, $\eta_{01}=\eta_{10}=0=g_{-+}\left({e_{0}}^{+}{e_{1}}^{-}+{e_{0}}^{-}{e_{1}}^{+}\right)$. We choose ${e_{0}}^{+}={e_{0}}^{-}={e_{1}}^{-}=-e^{-\rho(x^+,x^-)}$, and ${e_{1}}^{+}=e^{-\rho(x^+,x^-)}$. }
\begin{eqnarray}\lb{cg}
2\partial_{+}\Psi_{1}-\frac{\Psi_{1}}{\left(x^{+}-x^{-}\right)}=0, \quad -2\partial_{-}\Psi_{2}-\frac{\Psi_{2}}{\left(x^{+}-x^{-}\right)}=0.
\end{eqnarray}

The wave function in JT gravity spacetime must satisfy the following two boundary conditions: the wave function is zero at the AdS$_{2}$ spacetime boundary and it is finite at the past event horizon or the future event horizon of the extreme black hole in JT gravity. Combine the two boundary conditions and the eq.(\ref{cg}), we can find a particular solution of the wave function distribution beyond the extremal black hole horizon:
\begin{eqnarray}\lb{1}
\Psi_{1}(x^{+},x^{-})&=&\frac{1}{\sqrt{x^{-}}}\left(x^{-}-x^{+}\right)^{\frac{1}{2}}+i\frac{1}{\sqrt{x^{-}}}\left(x^{-}-x^{+}\right)^{\frac{1}{2}},\nonumber\\
\Psi_{2}(x^{+},x^{-})&=&\frac{1}{\sqrt{-x^{+}}}\left(x^{-}-x^{+}\right)^{\frac{1}{2}}+i\frac{1}{\sqrt{-x^{+}}}\left(x^{-}-x^{+}\right)^{\frac{1}{2}}.
\end{eqnarray}

\subsection{The dilaton }
 In the conformal gauge, using the general metric in two dimensional gravity in eq.(\ref{388}), and from eq.(\ref{Equation}) we finally have \footnote{In conformal gauge $ds^2=-e^{2\rho(x^+,x^-)}dx^+ dx^-$, we use the the following identities to get the equations of motion.
\begin{eqnarray}
\sqrt{-g}=\frac{1}{2}e^{2\rho}, \; g_{+-}=g_{-+}=-\frac{1}{2}e^{2\rho}, \; \square=g^{\mu\nu}\nabla_{\mu}\nabla_{\nu}=-4e^{-2\rho}\partial_{+}\partial_{-},\nonumber\\
\nabla_{+}\nabla_{+}\phi=\partial_{+}\partial_{+}\phi-2\partial_{+} \rho \partial_{+}\phi
 ,\; \nabla_{-}\nabla_{-}\phi=\partial_{-}\partial_{-}\phi-2\partial_{-} \rho \partial_{-} \phi, \; \nabla_{+}\nabla_{-}\phi=\partial_{+}\partial_{-}\phi.
\end{eqnarray}},\;

\begin{eqnarray}
&&\text{(1) For\ the\ metric}\ g_{+-}: \frac{e^{2\rho}}{2}\left(\phi_0-\phi\right)-\partial_{+}\partial_{-}\phi=\frac{i\alpha\kappa^{2}}{8}
\overline{\Psi}\left(\gamma_{-}\overrightarrow{D_{+}}-\overleftarrow{D_{+}}\gamma_{-}+\gamma_{+}\overrightarrow{D_{-}}-\overleftarrow{D_{-}}\gamma_{+}\right)\Psi, \label{guo}\\
&&\text{(2) For\ the\ metric}\ g_{++}: \partial_{+}\partial_{+}\phi-2\partial_{+} \rho \partial_{+}\phi=\frac{i\alpha\kappa^{2}}{4}\overline{\Psi}\left(\gamma_{+}\overrightarrow{D_{+}}-\overleftarrow{D_{+}}\gamma_{+}\right)\Psi, \label{chang}\\
&&\text{(3) For\ the\ metric}\ g_{--}:\partial_{-}\partial_{-}\phi-2\partial_{-} \rho \partial_{-} \phi=\frac{i\alpha\kappa^{2}}{4}\overline{\Psi}\left(\gamma_{-}\overrightarrow{D_{-}}-\overleftarrow{D_{-}}\gamma_{-}\right)\Psi. \label{zhong}
\end{eqnarray}
 The direction of the tetrad can be arbitrarily selected, we choose ${e_{0}}^{+}={e_{0}}^{-}={e_{1}}^{-}=-e^{-\rho}$ and ${e_{1}}^{+}=e^{-\rho}$. Then we can get the expression for the connection $\Gamma_{\mu}$ and the matrix $\gamma_{\mu}$ in the conformal gauge:
 \begin{eqnarray}
\gamma_{+}=\frac{e^{\rho}}{2}\left(\gamma^0+\gamma^1\right),\;\gamma_{-}=\frac{e^{\rho}}{2}\left(\gamma^0-\gamma^1\right), \label{guog}\\
\Gamma_{+}=\frac{\partial_{+}\rho}{2}\gamma^0\gamma^1,\;\Gamma_{-}=\frac{\partial_{-}\rho}{2}\gamma^1\gamma^0. \label{guogg}
\end{eqnarray}
Next, we substitute the 2-component massless Dirac spinor (\ref{gcz}) into the right hand side of eq.(\ref{guo}), eq.(\ref{chang}) and eq.(\ref{zhong}). Using eq.(\ref{guog}) and eq.(\ref{guogg}), then we have
\begin{eqnarray}
&\overline{\Psi}\left(\gamma_{-}\overrightarrow{D_{+}}-\overleftarrow{D_{+}}\gamma_{-}+\gamma_{+}\overrightarrow{D_{-}}-\overleftarrow{D_{-}}\gamma_{+}\right)\Psi=0,&\\
&\overline{\Psi}\left(\gamma_{+}\overrightarrow{D_{+}}-\overleftarrow{D_{+}}\gamma_{+}\right)\Psi=\frac{e^\rho}{2} \left(-2\Psi_{2}^{\ast}\partial_{+}\Psi_{2}+2\Psi_{2}\partial_{+}\Psi_{2}^{\ast}\right)&,\label{ganw}\\
&\overline{\Psi}\left(\gamma_{-}\overrightarrow{D_{-}}-\overleftarrow{D_{-}}\gamma_{-}\right)\Psi=\frac{e^\rho}{2} \left(-2\Psi_{1}^{\ast}\partial_{-}\Psi_{1}+2\Psi_{1}\partial_{-}\Psi_{1}^{\ast}\right)&.\label{ganwc}
\end{eqnarray}
Substituting the particular solution of the 2-component massless Dirac spinor (\ref{1}) back into the the right hand side of eq.(\ref{ganw}) and eq.(\ref{ganwc}), we can find
\begin{eqnarray}
-2\Psi_{2}^{\ast}\partial_{+}\Psi_{2}+2\Psi_{2}\partial_{+}\Psi_{2}^{\ast}=0, \; -2\Psi_{1}^{\ast}\partial_{-}\Psi_{1}+2\Psi_{1}\partial_{-}\Psi_{1}^{\ast}=0.
\end{eqnarray}
Finally, the equation of motion for the dilaton becomes
\begin{eqnarray}\lb{2}
\frac{2}{\left(x^{+}-x^{-}\right)^{2}}\left(\phi_0-\phi\right)-\partial_{+}\partial_{-}\phi&=&0,\nonumber\\
\frac{2}{\left(x^{+}-x^{-}\right)}\partial_{+}\left(\frac{\left(x^{+}-x^{-}\right)^{2}}{4}\partial_{+}\phi\right)&=&0,\nonumber\\
\frac{2}{\left(x^{+}-x^{-}\right)}\partial_{-}\left(\frac{\left(x^{+}-x^{-}\right)^{2}}{4}\partial_{-}\phi\right)&=&0.
\end{eqnarray}
We can solve the equation for the dilaton
\bqn
\phi=\phi_0+\frac{a+b\left(x^{+}+x^{-}\right)+cx^{+}x^{-}}{\left(x^{+}-x^{-}\right)},
\eqn
where $a$,$b$ and $c$ are constants which determine the dilaton of JT gravity.

In particular, the dilaton diverges at the  conformal boundary, and the location of this physical boundary is imposed by the boundary condition \cite{Maldacena:2016upp}:
\begin{eqnarray}
g_{uu}\mid_{bdy}=\frac{1}{\varepsilon^2}, \quad \phi=\phi_{b}=\frac{\phi_{r}}{\varepsilon}+\phi_0,
\end{eqnarray}
where $u$ is the physical boundary time, with $\varepsilon$ the UV cutoff.

 The metric in JT gravity has $SL(2,R)$ isometry. For the extreme black hole in JT gravity,  under the $SL(2,R)$ transformation the dilaton profiles can be recast as
\bqn
\phi=\phi_0+\frac{2\phi_{r}}{\left(x^{+}-x^{-}\right)}.
\eqn

\section{ The two point correlators}\lb{4d-far}
\renewcommand{\theequation}{3.\arabic{equation}}\setcounter{equation}{0}

\subsection{The primary fermion field correlator in two dimensional Minkowski spacetime}

We consider a free Dirac field in two dimensions, it satisfies the Dirac equation and the canonical anticommutation relations :
\begin{eqnarray}
\left(i\gamma^{\mu}\partial_{\mu}-m\right)\Psi=0, \quad \{\Psi_{\alpha}(\vec{x}),\Psi_{\beta}^{\dag}(\vec{y})\}=\delta_{\alpha\beta}\delta(\vec{x}-\vec{y}),
\end{eqnarray}
where $x$ and $y$ lie on the Cauchy surface with $t$ = constant. And the two point field correlator in two dimensional Minkowski spacetime is \footnote{Casini first used $\Psi^{\dagger}$ instead of $\bar{\Psi}$ in their computation for the two point field correlator in \cite{Casini:2009vk,Casini:2009sr}. There exists local Lorentz boost transformations in $2D$ spacetime, for which $\bar{\Psi}\Psi$ is an invariant for fermions, the vacuum expectation value of $\Psi\bar{\Psi}$ called Feynman propagator is defined as $\langle0|\Psi(\vec{x})\bar{\Psi}(\vec{y})|0\rangle$ in QFT. In fact, Casini defined the two point field correlator as $C(\vec{x},\vec{y})=\langle0|\Psi(\vec{x})\Psi^{\dag}(\vec{y})|0\rangle$ in order to calculate the entanglement entropy of a massless Dirac field with the correlator trace formula (\ref{357}). }:
\bqn\lb{co}
C(\vec{x},\vec{y})=\langle0|\Psi(\vec{x})\Psi^{\dag}(\vec{y})|0\rangle=\int\frac{dp}{2\pi}\frac{\left(p_{\mu}\gamma^{\mu}+m\right)}{2\sqrt{p^2+m^2}}\gamma^{0}e^{-ip\cdot\left(x-y\right)}.
\eqn
The integral of the two point field correlator in eq.(\ref{co})is \cite{Casini:2009vk}:

\bqn
C(x,y)=\frac{1}{2}\delta\left(x-y\right)\textbf{1}+\frac{m}{2\pi}K_{0}\left(m|x-y|\right)\gamma^{0}+\frac{im}{2\pi}K_{1}\left(m\left(x-y\right)\right)\gamma^{0}\gamma^{1},
\eqn
where $K_{n}(x)$ is the standard modified Bessel function, and in the massless limit this gives the two point correlator for the primary fermion field in two dimensional flat spacetime:
\bqn\lb{cor}
C(x,y)=\frac{1}{2}\delta\left(x-y\right)\textbf{1}+\frac{i}{2\pi}\frac{1}{\left(x-y\right)}\gamma^{0}\gamma^{1}.
\eqn

\subsection{The primary fermion field correlator in JT gravity}

In general, the metric in 2D conformally flat spacetime is:
\bqn
ds^2=-e^{2\rho(x^+,x^-)}dx^+ dx^-=-\Omega^{-2}(x^+,x^-)dx^+ dx^-,
\eqn
where $\Omega=e^{-\rho}$ is the conformal factor. Two dimensional JT gravity is locally AdS$_{2}$ spacetime, the conformal factor is $\Omega=\left(x^{+}-x^{-}\right)/2$.

In CFT  method, the two point correlation function for primary operators on a curved manifold with Weyl rescaled metric $\Omega^{-2}g$ in terms of those with metric $g$ satisfies the following transformation relation under Weyl transformations \cite{Almheiri:2019psf,DiFrancesco:1997nk}:
\bqn\lb{386}
\left\langle\Phi\left(x_{1}, \bar{x}_{1}\right) \tilde{\Phi}\left(x_{2}, \bar{x}_{2}\right)\right\rangle_{\Omega^{-2}g}=\Omega\left(x_{1}, \bar{x}_{1}\right)^{\Delta}\Omega\left(x_{2},\bar{x}_{2}\right)^{\Delta}\left\langle\Phi\left(x_{1},\bar{x}_{1}\right)\tilde{\Phi}\left(x_{2}, \bar{x}_{2}\right)\right\rangle_{g},
\eqn
where $\Delta$ is the scale dimension for the twist field and $\left\langle\Phi\left(x_{1},\bar{x}_{1}\right)\tilde{\Phi}\left(x_{2}\bar{x}_{2}\right)\right\rangle_{g}$ is the two point correlation function for primary operators in two dimensional flat spacetime.

The free massless fermion field is also the primary field with the scale dimension $\Delta = 1/2$. Combining eq.(\ref{cor}) and eq.(\ref{386}), then we can get the two point correlators of primary fermion fields $C(x,y)_{\Omega^{-2}g}$ in JT gravity after Weyl transformed from $ds^2 = -dx^+ dx^-$ to $ds^2 = -\Omega^{-2}(x^+,x^-)dx^+ dx^-$:
\begin{eqnarray}\lb{rejt}
C(x,y)_{g}&=&\left\langle\Phi\left(x,\bar{x}\right)\tilde{\Phi}\left(y,\bar{y}\right)\right\rangle_{g}=\frac{1}{2}\delta\left(x-y\right)\textbf{1}+\frac{i}{2\pi}\frac{1}{\left(x-y\right)}\gamma^{0}\gamma^{1}\nonumber\\
\Longrightarrow C(x,y)_{\Omega^{-2}g}&=&\left\langle\Phi\left(x, \bar{x}\right) \tilde{\Phi}\left(y, \bar{y}\right)\right\rangle_{\Omega^{-2}g}=\frac{\left(xy\right)^{\frac{1}{2}}}{2}\delta\left(x-y\right)\textbf{1}+\frac{i}{2\pi}\frac{\left(xy\right)^{\frac{1}{2}}}{\left(x-y\right)}\gamma^{0}\gamma^{1}.
\end{eqnarray}

\section{Entanglement entropy}
\begin{figure}[h]
\includegraphics[height=4.7cm]{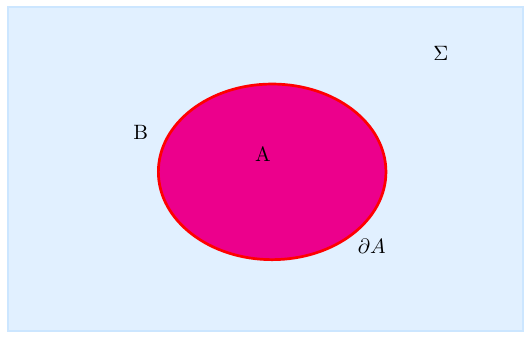}
\caption{A continuum QFT has been spatially divided into two components on a Cauchy slice $\Sigma$. The region $B$ is the complement of the region $A$, and the red curve $\partial A$ is the entangling surface which is a spacetime codimension-2 surface.}
\label{ee}
\end{figure}

\lb{4d-near}
 \renewcommand{\theequation}{4.\arabic{equation}}\setcounter{equation}{0}
The entanglement entropy (von-Neumann entropy) provides us with a convenient way to measure the degree of entanglement between two quantum systems in QFT. We choose the total quantum system as a pure quantum state with the density matrix $\rho=|\Psi\rangle\langle\Psi|$. The reduced density matrix for the subsystem $A$ is $\rho_{A}=Tr_{B}|\Psi\rangle\langle\Psi|$, which is obtained by taking a partial trace over the subsystem $B$ of the total density matrix (see the Fig.\ref{ee}). The entanglement entropy  for the subsystem $A$ is the corresponding von Neumann entropy:
\bqn\lb{1812}
S_{A}=-Tr\left(\rho_{A}\ln\rho_{A}\right).
\eqn

For the 1+1 dimensional quantum system at criticality, the continuum limit is a conformal field theory with central charge $c$. The renormalized  entanglement entropy of a single interval in vacuum state in flat spacetime can be calculated by Cardy formula \cite{Calabrese:2004eu,Calabrese:2009qy}:
\bqn
S=\frac{c}{3}\log\ell,
\eqn
where $\ell$ is the length of the single interval on the line in vacuum. After Weyl transformed from $ds^2 = -dx^+ dx^-$ to $ds^2 = -\Omega^{-2}(x^+,x^-)dx^+ dx^-$, entanglement entropy \textcolor{red}{in} 2D conformally flat spacetime is transformed as \cite{Almheiri:2019psf,Gan:2022jay}:
\bqn
S_{\Omega^{-2} g}=S_{g}-\frac{c}{6}\sum_{endpoints}\log(\Omega)=S_{g}+\frac{c}{6}\sum_{endpoints}\log(e^{\rho}).
\eqn

Entanglement entropy is related to the reduced density matrix of the region $V$, the problem of finding an explicit expression for the local density matrix $\rho_{V}$ is equivalent to solving the resolvent of the two point correlators $C_{V}$ in the massless case. {\it Resolvent} is a standard technique in complex analysis, the use of the resolvent technique for free massless fermions was first introduced in \cite{Casini:2009vk} to study the entanglement entropy in vacuum on the plane, and subsequently for the entanglement entropy of a chiral fermion on the torus \cite{Blanco:2019cet,Fries:2019acy,He:2022xkh}. In this section we will first review the derivation of the entanglement entropy for a massless Dirac field in two dimensional vacuum Minkowski spacetime in terms of the resolvent technique, and we can get the entanglement entropy of a single interval for a massless Dirac field in 2D conformally flat JT gravity by redefining the field in terms of the conformal factor as the Fermion field.

\subsection{Entanglement entropy for a massless Dirac field in two dimensional vacuum Minkowski spacetime}

The two point function $C_{V}$ is related to the reduced density matrix of the region $V$ by the condition:
\bqn
C_{V}(x,y)=\langle\Psi(x)\Psi^{\dag}(y)\rangle=Tr\left(\rho_{V}\Psi(x)\Psi^{\dag}(y)\right).
\eqn
Then the expression for the entanglement entropy of the region $V$ can be given by a propagator trace formula (see Appendix (\ref{app-4})) \cite{Casini:2009vk,Casini:2009sr,Casini:2007bt}:
\bqn\lb{357}
S_{V}=-Tr[\left(1-C_{V}\right)\log\left(1-C_{V}\right)+C_{V}\log C_{V}].
\eqn
The {\it resolvent} of the two point function $C_{V}$ is defined as:
\bqn\lb{re}
R_{V}(\xi):=\left(C_{V}+\xi-1/2\right)^{-1}.
\eqn
Combining the the expression for the resolvent (\ref{re}), the entanglement entropy can be rewritten as:
\bqn\lb{res}
S_{V}=-Tr\int^{+\infty}_{1/2}d\xi[(\xi-1/2)[R(\xi)-R(-\xi)]-\frac{2\xi}{\xi+1/2}].
\eqn

In eq.(\ref{re}), the inverse of an operator for the propagator is understood in the sense of a kernel that satisfies the following equation:
\bqn\lb{ke}
\int_{V}dz R_{V}(\xi;x,z)R_{V}^{-1}(\xi;z,y)=\delta\left(x-y\right)=\int_{V}dz R_{V}(\xi;x,z)[C(z,y)+\left(\xi-1/2\right)\delta(z,y)].
\eqn
Substituting (\ref{cor}) into (\ref{ke}) one obtains a singular integral equation \cite{reso}:
\bqn\lb{sie}
\xi R_{V}(x,y)-\frac{i}{2\pi }\int_{V}\frac{R_{V}(x,z)}{z-y}dz=\delta\left(x-y\right).
\eqn
 Fortunately, we can solve the resolvent for this integral operator inside a region formed by $n$ disjoint intervals $(u_{i},v_{i})$ by the Plemelj formulae \cite{reso} in the theory of singular integral equations (see Appendix (\ref{app-a})). The resolvent of the two point function $C_{V}$ (see Appendix (\ref{apx.b})):
\bqn\lb{resol}
R_{V}(\xi) =\left(\xi^2-1/4 \right)^{-1}
\left(\xi\,\delta(x-y)\,+\frac{i }{2\pi}   \frac{e^{-\frac{i}{2\pi} \log\left(\frac{\xi-1/2}{\xi+1/2}\right)\, \left(z(x)-z(y)\right) }}{x-y}
\right),
\eqn
where the function $z(x)$ is
\bqn
z(x)=\log\left(-\frac{\prod_{i=1}^n (x-u_i)}{\prod_{i=1}^n (x-v_i)}\right).
\eqn

Substituting (\ref{resol}) into (\ref{res}), then we have
\bqn
S_{V}=-\frac{2}{\pi}\int^\infty_{1/2} d\xi\, \int_V dx\, \lim_{y\rightarrow x}  \frac{\sin\left[ \frac{1}{2\pi} \log\left(\frac{\xi-1/2}{\xi+1/2}\right)\, \left(z(x)-z(y)\right) \right]}{(\xi+1/2)\,(x-y)}\,.
\eqn
Integrating over $\xi$ first, we can get the entanglement entropy in $n$ disjoint intervals for a massless Dirac field in two dimensional vacuum Minkowski spacetime:
\begin{eqnarray}
S_{V}&=&2\int_V dx\, \lim_{y\rightarrow x}  \frac{\frac{z(x)-z(y)}{2}\coth\left(\left(z(x)-z(y)\right)/2\right)-1}{(x-y)\left(z(x)-z(y)\right)}=\frac{1}{6}\int_V dx\,\sum_{i=1}^n \left(\frac{1}{x-u_i}-\frac{1}{x-v_i}\right)\nonumber\\
&\,&\hspace{1cm}=\frac{1}{3} \left( \sum_{i,j}\log|v_i-u_i|-\sum_{i<j} \log|u_i-u_j| -\sum_{i<j} \log|v_i-v_j|-n \log \epsilon \right)\,,\label{sv}
\end{eqnarray}
where $\epsilon$ is a distance cutoff introduced in the last integration, and the Virasoro central charge of the primary fermion field is $c=1$. For a single interval in 2D vacuum flat spacetime on the plane, we can verify the Cardy formula for the renormalized entanglement entropy $S=\frac{c}{3}\log\ell$.

\subsection{Entanglement entropy  for a massless Dirac field in JT gravity}

In this subsection, we apply the resolvent technique to 2D conformally flat spacetime. We begin by redefining the field in terms of the conformal factor as the Fermion field \footnote{We would like to thank Yiming Chen for bringing this point to our attention.}.
Let us consider the rescaling field, which is given by:
\bqn
\hat\Psi(\vec{x})=\Omega^\Delta(\vec{x}) \Psi(\vec{x})=\Omega^\frac{1}{2}(\vec{x}) \Psi(\vec{x}),
\eqn
Using this rescaling field, we can use the same approach as described in the previous subsection and obtain the same results as shown in equation \eqref{sv}.
After performing the calculations using the original field $\Psi(\vec{x})$, one finally \textcolor{red}{finds}
\begin{eqnarray}
S_{V}=\frac{1}{3} \left( \sum_{i,j}\log|\frac{v_i-u_i}{(u_iv_i)^{1/2}}|-\sum_{i<j} \log|\frac{u_i-u_j}{(u_iu_j)^{1/2}}| -\sum_{i<j} \log|\frac{v_i-v_j}{(v_iv_j)^{1/2}}|-n \log \epsilon \right)\,,
\end{eqnarray}

The renormalized entanglement entropy for a massless Dirac field of a single interval in JT gravity is \footnote{In $2D$ dilaton gravity, the generalized entropy of Hawking radiation is given by
\bqn\nonumber
S_{\text{gen}}(R)=S_{\text{gravity}}+S_{\text{matter}}=\frac{\phi}{4G_N}+\frac{c}{6}\log\frac{\ell^2}{\epsilon^2\cdot(\Omega_{A}\Omega_{B})}.
\eqn
We can omit the UV cutoff parameter $\epsilon$, since it can be absorbed in the renormalization of Newton constant $G_N$ \cite{Almheiri:2019psf, Gan:2022jay}. Then the renormalized entanglement entropy for a massless Dirac field of a single interval in JT gravity can be given by eq.(\ref{eejt}).}:
\bqn\lb{eejt}
S=\frac{1}{6}\log\frac{\ell^2}{\Omega_{A}\Omega_{B}}=\frac{1}{3}\log\frac{|x-y|}{(xy)^{\frac{1}{2}}},
\eqn
where the Virasoro central charge of the massless Dirac field is $c = 1$.

\section{Conclusion and discussion}\lb{discussion}
\renewcommand{\theequation}{5.\arabic{equation}}\setcounter{equation}{0}
In this paper we get the particular solution of the wave function outside the extremal black hole horizon in JT gravity, it is very important for the follow-up study of extracting extremal black hole information with modular flow in JT gravity. The specific expression for the modular flow of 2D free massless fermion depends on the wave function, other papers derived the modular flow formula for 2D free massless fermions but didn't give us the specific expression for the wave function \cite{Chen:2019iro,Blanco:2019cet,Erdmenger:2020nop,Reyes:2021npy}.

In CFT$_2$ methods, a convenient way to compute entropies of intervals is using the {\it replica trick} to compute the R\'{e}nyi entropy for integer index $n$:
\bqn
S_{n}(V)=\frac{1}{1-n}\log Tr\rho^n_{V}.
\eqn
Taking the limit $n\rightarrow1$, we can derive the entanglement entropy of the primary fermion fields\cite{Almheiri:2019psf,Calabrese:2004eu,Calabrese:2009qy}.
There is a simpler technique called {\it resolvent} to derive the entanglement entropy for 2D free massless fermions in comparison to the CFT method called replica trick. In this paper we calculate the two point correlators of primary fermion fields in JT gravity under Weyl transformations. And we redefine the fields in terms of the conformal factor as the fermion fields, then we use the resolvent technique as described in two dimensional vacuum Minkowski spacetime to derive the renormalized entanglement entropy for  massless Dirac fields in JT gravity.

In this work, we have calculated the wave function and derived the entanglement entropy for the primary fermion fields outside the extremal black hole horizon in JT gravity. In this case, we only consider the quantum entanglement between the free massless fermions outside the extremal black hole horizon. In the future study, we will go a step further by paying attention to the following points:\\
$(a)$ For the entanglement between the free massless fermions inside the horizon and outside the horizon, we should regard the whole spacetime as a total quantum entanglement system composed of the extremal black hole and Hawking radiation outside the horizon. The degrees of freedom for the free massless fermions located inside the horizon represent the degrees of freedom for the extremal black hole, and the degrees of freedom for the free massless fermions located outside the horizon represent the degrees of freedom for Hawking radiation particles.\\
$(b)$ In order to calculate the entanglement entropy for the free massless fermions inside the horizon and outside the horizon, we should consider the entanglement island inside the extremal black hole interior in JT gravity. We may calculate the fine grained entropy of the extremal black hole and Hawking radiation via the semiclassical method called \textit{island rule}.

\section*{Acknowledgments}

We thank Hong-An Zeng for helpful discussions on the resolvent of the primary fermion correlator in 2D vacuum Minkowski spacetime. This work is supported by the National Natural Science Foundation of China with Grant No. 11975116.

\begin{appendix}
\section{The equations of motions in the background of JT gravity coupled to primary fermion fields}\lb{app-1}
 \renewcommand{\theequation}{A.\arabic{equation}}\setcounter{equation}{0}
The total action functional for JT gravity coupled to primary fermions is eq.(\ref{st}), we can get the classical equation of motion by varying the metric of the total action:
\begin{eqnarray}\lb{zfc}
\frac{\delta S}{\delta g^{\mu\nu}}=0, \; \Longrightarrow -\frac{\delta S_{JT}}{\delta g^{\mu\nu}} = \frac{\alpha\delta S_{D}}{\delta g^{\mu\nu}}.
\end{eqnarray}

The variation of (\ref{Dirac}) with respect to the frame vector indices $e^{a\mu}$ is \cite{Freedman:2012zz}:
\bqn\lb{sdb}
\delta S_{D}=\int d^2x\frac{i}{4}\sqrt{-g}\overline{\Psi}\left[\gamma_{a}\overleftrightarrow{D_{\mu}}+\gamma_{\mu}{e_{a}}^{\rho}\overleftrightarrow{D_{\rho}}\right]\Psi\delta e^{a\mu},
\eqn
where $\eta^{ab}{e_b}^{\mu}=e^{a\mu}$. We use $\delta e^{a\mu}=\frac{1}{4}{e^a}_{\nu}\delta g^{\mu\nu}$. By the variation of the metric $g^{\mu\nu}$, then eq.(\ref{sdb}) can be written as:
\bqn\lb{sddg}
\delta S_{D}=\int d^2x\frac{i}{16}\sqrt{-g}\overline{\Psi}\left[\gamma_{\nu}\overleftrightarrow{D_{\mu}}+\gamma_{\mu}\overleftrightarrow{D_{\nu}}\right]\Psi\delta g^{\mu\nu},
\eqn
where we have used the following contractions in (\ref{sddg}),
\begin{eqnarray}
\gamma_{a}{e^a}_{\nu}=\gamma_{\nu}, \;{e_a}^{\rho}{e^a}_{\nu}=\delta^{\rho}_{\nu}.
\end{eqnarray}

For the classical bulk term action of JT gravity(\ref{bjt}), using the standard relations\cite{Carroll:2004st},
\begin{eqnarray}
\delta\sqrt{-g}=-\frac{1}{2}\sqrt{-g}g_{\mu\nu}\delta g^{\mu\nu}, \;\phi g^{\mu\nu}\delta R_{\mu\nu}=-\left[\left(\nabla_{\mu}\nabla_{\nu}-g_{\mu\nu}\square\right)\phi\right]\delta g^{\mu\nu}.
\end{eqnarray}
By varying the metric $g^{\mu\nu}$ in 2D spacetime, we get:
\begin{eqnarray}\lb{jtbf}
\delta S_{JT}&=&\frac{1}{16\pi G_N}\int d^2x\left[\delta(\sqrt{-g})\left(\phi R+2\phi-2\phi_0\right)+\sqrt{-g}\phi\delta(g^{\mu\nu}R_{\mu\nu})\right]\nonumber\\
&=&\frac{1}{16\pi G_N}\int d^2x\sqrt{-g}\left[-\frac{1}{2}\sqrt{-g}g_{\mu\nu}\delta g^{\mu\nu}\left(\phi R+2\phi-2\phi_0\right)+\sqrt{-g}\phi R_{\mu\nu}\delta g^{\mu\nu}+\sqrt{-g}\phi g^{\mu\nu}\delta R_{\mu\nu}\right]\nonumber\\
&=&\frac{1}{16\pi G_N}\int d^2x\sqrt{-g}\left[-\frac{1}{2}\sqrt{-g}g_{\mu\nu}\delta g^{\mu\nu}\left(\phi R+2\phi-2\phi_0\right)+\sqrt{-g}\phi R_{\mu\nu}\delta g^{\mu\nu}+\sqrt{-g}\left[g_{\mu\nu}\square-\nabla_{\mu}\nabla_{\nu}\right]\phi\delta g^{\mu\nu}\right]\nonumber\\
&=&\frac{1}{32\pi G_N}\int d^2x\sqrt{-g}\left[2g_{\mu\nu}\left(\phi_0-\phi\right)+2\phi\left(R_{\mu\nu}-\frac{1}{2}g_{\mu\nu}R\right)+2g_{\mu\nu}\square\phi-2\nabla_{\mu}\nabla_{\nu}\phi\right]\delta g^{\mu\nu}.
\end{eqnarray}
In 2D gravity, we can easily calculate that the Einstein tensor is zero. In the last term in eq.(\ref{jtbf}), we have $G_{\mu\nu}=R_{\mu\nu}-\frac{1}{2}g_{\mu\nu}R=0$. Then eq.(\ref{jtbf}) becomes
\bqn\lb{ljtbf}
\delta S_{JT}=\frac{1}{32\pi G_N}\int d^2x\sqrt{-g}\left[2g_{\mu\nu}\left(\phi_0-\phi\right)+2g_{\mu\nu}\square\phi-2\nabla_{\mu}\nabla_{\nu}\phi\right]\delta g^{\mu\nu}.
\eqn

Finally, substituting (\ref{sddg}) and (\ref{ljtbf}) into (\ref{zfc}), then we get the classical equation of motion in JT gravity coupled to primary fermion fields:
\bqn
g_{\mu \nu}\left(\phi-\phi_0\right)+\nabla_{\mu}\nabla_{\nu}\phi-g_{\mu \nu}\square\phi=\frac{i\alpha\kappa^{2}}{8}\overline{\Psi}\left(\gamma_{\nu}\overleftrightarrow{D_{\mu}}+\gamma_{\mu}\overleftrightarrow{D_{\nu}}\right)\Psi.
\eqn

\section{Singular integral equations and the Plemelj formulae}
\begin{figure}[h]
\includegraphics[height=4.7cm]{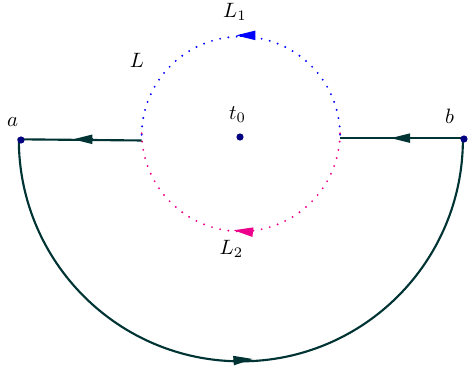}
\caption{$L$ is the line segment with two endpoints $a$ and $b$, $t_0$ is the midpoint of the line segment $L$. $L_1$ is the blue semicircle which is going in the counterclockwise direction, and $L_2$ is the red semicircle which is going in the clockwise direction. $L_1+L$ represents the contour that contains $t_0$, $L_2+L$ represents the contour that doesn't contain $t_0$. $L_1-L_2$ represents a complete circle which is going in the counterclockwise direction. }
\label{wedao}
\end{figure}

\lb{app-a}
\renewcommand{\theequation}{B.\arabic{equation}}\setcounter{equation}{0}
For the entire complex plane (see the Fig.\ref{wedao}), we can get the integral formula of the function $\varphi(t_0)$ by the Cauchy's integral formula \cite{reso}:
\bqn\lb{b1}
\varphi(t_0)=\frac{1}{2\pi i}\oint_{L_1-L_2}\frac{\varphi(t)dt}{t-t_0}=\frac{1}{2\pi i}\int_{L_1}\frac{\varphi(t)dt}{t-t_0}-\frac{1}{2\pi i}\int_{L_2}\frac{\varphi(t)dt}{t-t_0}.
\eqn
From the eq.(\ref{b1}) we can easily see:
\begin{eqnarray}
\frac{1}{2\pi i}\int_{L_1}\frac{\varphi(t)dt}{t-t_0}=\frac{1}{2}\varphi(t_0), \; \frac{1}{2\pi i}\int_{L_2}\frac{\varphi(t)dt}{t-t_0}=-\frac{1}{2}\varphi(t_0).
\end{eqnarray}

Equations of the type
\bqn\lb{b3}
A(t_0)\varphi(t_0)+\frac{B(t_0)}{\pi i}\int_{L}\frac{\varphi(t)dt}{t-t_0}=f(t_0)
\eqn
is called singular integral equations. We define the following functions:
\begin{eqnarray}\lb{b4}
\Phi(t_0)&\equiv &\frac{1}{2\pi i}\int_{L}\frac{\varphi(t)dt}{t-t_0}\\
\Phi^+(t_0)&\equiv &\frac{1}{2\pi i}\int_{L_1}\frac{\varphi(t)dt}{t-t_0}+\frac{1}{2\pi i}\int_{L}\frac{\varphi(t)dt}{t-t_0}=\frac{1}{2}\varphi(t_0)+\frac{1}{2\pi i}\int_{L}\frac{\varphi(t)dt}{t-t_0}\\
\Phi^-(t_0)&\equiv &\frac{1}{2\pi i}\int_{L_2}\frac{\varphi(t)dt}{t-t_0}+\frac{1}{2\pi i}\int_{L}\frac{\varphi(t)dt}{t-t_0}=-\frac{1}{2}\varphi(t_0)+\frac{1}{2\pi i}\int_{L}\frac{\varphi(t)dt}{t-t_0}.
\end{eqnarray}
Substituting eq.(\ref{b4}) into eq.(\ref{b3}), then we have
\begin{eqnarray}\lb{b5}
\left(A(t_0)+B(t_0)\right)\Phi^+(t_0)-\left(A(t_0)-B(t_0)\right)\Phi^-(t_0)=f(t_0)\\
\Longrightarrow \Phi^+(t_0)=\frac{A(t_0)-B(t_0)}{A(t_0)+B(t_0)}\Phi^-(t_0)+\frac{f(t_0)}{A(t_0)+B(t_0)} .
\end{eqnarray}
We define $G(t_0)\equiv \frac{A(t_0)-B(t_0)}{A(t_0)+B(t_0)}$ and $g(t_0)=\frac{f(t_0)}{A(t_0)+B(t_0)}$, then eq.(\ref{b5}) is reduced to a simpler singular integral equation:
\bqn\lb{b8}
\Phi^+(t_0)=G(t_0)\Phi^-(t_0)+g(t_0).
\eqn

We define a homogeneous equation :
\begin{eqnarray}\lb{aabb}
X^+(t_0)=G(t_0)X^-(t_0), \; G(t_0)=\frac{X^+(t_0)}{X^-(t_0)}=\frac{A(t_0)-B(t_0)}{A(t_0)+B(t_0)}.
\end{eqnarray}
By taking logarithms, we obtain
\bqn\lb{pre}
\log X^+(t_0)-\log X^-(t_0)=\log G(t_0),
\eqn
where eq.(\ref{pre}) is the Plemelj formulae with the corresponding solution \cite{reso}:
\begin{eqnarray}
\log X(t_0)=\frac{1}{2\pi i}\int_{L}\frac{\log G(t)dt}{t-t_0}, \; \log X^{\pm}(t_0)=\pm\frac{1}{2}\log G(t_0)+\frac{1}{2\pi i}\int_{L}\frac{\log G(t)dt}{t-t_0}.
\end{eqnarray}
And the solution to $X^{\pm}(t_0)$ is
\bqn
X^{\pm}(t_0)=e^{\pm\frac{1}{2}\log G(t_0)+\frac{1}{2\pi i}\int_L\frac{\log G(t)dt}{t-t_0}}.
\eqn

Combining eq.(\ref{b8}) and eq.(\ref{aabb}), then we have
\bqn\lb{11}
\frac{\Phi^+(t_0)}{X^+(t_0)}-\frac{\Phi^-(t_0)}{X^-(t_0)}=\frac{g(t_0)}{X^+(t_0)}.
\eqn
Eq.(\ref{11}) is also the Plemelj formulae, and the corresponding solution is
\begin{eqnarray}\lb{55}
\frac{\Phi^+(t_0)}{X^+(t_0)}&= &\frac{1}{2}\frac{g(t_0)}{X^+(t_0)}+\frac{1}{2\pi i}\int_{L}\frac{g(t)dt}{X^+(t)(t-t_0)}\nonumber\\
\frac{\Phi^-(t_0)}{X^-(t_0)}&= &-\frac{1}{2}\frac{g(t_0)}{X^+(t_0)}+\frac{1}{2\pi i}\int_{L}\frac{g(t)dt}{X^+(t)(t-t_0)}.
\end{eqnarray}
\section{The resolvent of the primary fermion correlator in two dimensional vacuum Minkowski spacetime}\lb{apx.b}
\renewcommand{\theequation}{C.\arabic{equation}}\setcounter{equation}{0}
To solve the singular integral equation of the resolvent (\ref{sie}), we define
\bqn
\Phi^{\pm}(x,y)=\pm\frac{1}{2}R(x,y)+\frac{1}{2\pi i}\int_{L}\frac{R(x,z)}{z-y}dz,
\eqn
then we have
\begin{eqnarray}
\Phi^{+}(x,y)-\Phi^{-}(x,y)=R(x,y), \; \Phi^{+}(x,y)+\Phi^{-}(x,y)=\frac{1}{\pi i}\int_{L}\frac{R(x,z)}{z-y}dz.
\end{eqnarray}
Then eq.(\ref{sie}) can be written as
\bqn\lb{22}
\left(\xi+\frac{1}{2}\right)\Phi^{+}(x,y)-\left(\xi-\frac{1}{2}\right)\Phi^{-}(x,y)=\delta\left(x-y\right).
\eqn
We define a homogeneous equation :
\bqn\lb{33}
X^+(x,y)=G(\xi)X^-(x,y), \; G(\xi)=\frac{\xi-\frac{1}{2}}{\xi+\frac{1}{2}}.
\eqn
By taking logarithms, we obtain
\bqn
\log X^+(x,y)-\log X^-(x,y)=\log G(\xi),
\eqn
with the corresponding solution:
\begin{eqnarray}
\log X(x,y)=\frac{1}{2\pi i}\int_{L}\frac{\log G(\xi)dz}{z-y}, \; \log X^{\pm}(x,y)=\pm\frac{1}{2}\log G(\xi)+\frac{1}{2\pi i }\int_{L}\frac{\log G(\xi)dz}{z-y}.
\end{eqnarray}
For a single interval $L=[a,b]$, the solution to $X^{\pm}(x,y)$ is
\bqn\lb{88}
X^{\pm}(x,y)=e^{\pm\frac{1}{2}\log G(\xi)+\frac{1}{2\pi i}\log G(\xi)\log\frac{b-y}{y-a}}.
\eqn

Combining eq.(\ref{22}) and eq.(\ref{33}), then we can get the Plemelj formulae:
\bqn\lb{44}
\frac{\Phi^+(x,y)}{X^+(x,y)}-\frac{\Phi^-(x,y)}{X^-(x,y)}=\frac{f(x,y)}{X^+(x,y)}, \; f(x,y)=\frac{\delta(x-y)}{\xi+\frac{1}{2}}.
\eqn
Combining the solution to the Plemelj formulae (\ref{55}), we can get the solution to (\ref{44}):
\begin{eqnarray}\lb{66}
\frac{\Phi^+(x,y)}{X^+(x,y)}&= &\frac{1}{2}\frac{f(x,y)}{X^+(x,y)}+\frac{1}{2\pi i}\int_{L}\frac{f(x,z)dz}{X^+(x,z)(z-y)}\nonumber\\
\frac{\Phi^-(x,y)}{X^-(x,y)}&= &-\frac{1}{2}\frac{f(x,y)}{X^+(x,y)}+\frac{1}{2\pi i}\int_{L}\frac{f(x,z)dz}{X^+(x,z)(z-y)}\\
\Longrightarrow\Phi^+(x,y)&= &\frac{1}{2}f(x,y)+\frac{1}{2\pi i}X^+(x,y)\int_{L}\frac{f(x,z)dz}{X^+(x,z)(z-y)}\nonumber\\
\Longrightarrow\Phi^-(x,y)&= &-\frac{1}{2}\frac{f(x,y)}{G(\xi)}+\frac{1}{2\pi i}X^-(x,y)\int_{L}\frac{f(x,z)dz}{X^+(x,z)(z-y)}.
\end{eqnarray}
Then we can get the solution to the resolvent $R(x,y)$ \footnote{In the last term in eq.(\ref{77}), we have used the selectivity of the function $\delta(x)$ :
\bqn
\int\delta(x-z)f(z)dz=f(x).
\eqn}:
\begin{eqnarray}\lb{77}
R(x,y)&=&\Phi^{+}(x,y)-\Phi^{-}(x,y)=\frac{\xi}{\xi-\frac{1}{2}}f(x,y)-\frac{1}{2\pi i }\frac{X^+(x,y)}{\xi-\frac{1}{2}}\int_{L}\frac{f(x,z)dz}{X^+(x,z)(z-y)}\nonumber\\
&=&\frac{\xi\delta(x-y)}{(\xi-\frac{1}{2})(\xi+\frac{1}{2})}-\frac{1}{2\pi i }\frac{X^+(x,y)}{(\xi-\frac{1}{2})(\xi+\frac{1}{2})}\int_{L}\frac{\delta(x-z)dz}{X^+(x,z)(z-y)}\nonumber\\
&=&\frac{\xi\delta(x-y)}{(\xi-\frac{1}{2})(\xi+\frac{1}{2})}-\frac{1}{2\pi i}\frac{X^+(x,y)}{(\xi-\frac{1}{2})(\xi+\frac{1}{2})(X^+(x,x))(x-y)}.
\end{eqnarray}
Substituting (\ref{88}) into (\ref{77}), then we can get the expression for the resolvent $R(x,y)$ of a single interval $L=[a,b]$:
\begin{eqnarray}\lb{99}
R(x,y)&=&\left(\xi^2-1/4\right)^{-1}\left
(\xi\delta(x-y)+\frac{i}{2\pi}\frac{e^{-\frac{i}{2\pi}\log\left(\frac{\xi-\frac{1}{2}}{\xi+\frac{1}{2}}\right)\left(\log\left(-\frac{(x-a)}{(x-b)}\right)-\log\left(-\frac{(y-a)}{(y-b)}\right)\right)}}{(x-y)}\right).
\end{eqnarray}
When $L$ contains n disjoint intervals, where $L=(a_1,b_1)\cup(a_2,b_2)\cup\ldots\cup(a_n,b_n)$, the resolvent of the primary fermion correlator in multicomponent subsets of the $L$ in two dimensional vacuum Minkowski spacetime can be written as
\bqn
R(x,y) =\left(\xi^2-1/4 \right)^{-1}
\left(\xi\,\delta(x-y)\,+\frac{i }{2\pi}   \frac{e^{-\frac{i}{2\pi} \log\left(\frac{\xi-1/2}{\xi+1/2}\right)\, (z(x)-z(y)) }}{x-y}
\right),
\eqn
where the function $z(x)$ is
\bqn
z(x)=\log\left(-\frac{\prod_{i=1}^n (x-u_i)}{\prod_{i=1}^n (x-v_i)}\right).
\eqn

\section{Entanglement entropy for primary fermion fields given by a correlator trace formula}\lb{app-4}
\renewcommand{\theequation}{D.\arabic{equation}}\setcounter{equation}{0}

The creation and annihilation operators $\Psi_i^{\dag}$ and $\Psi_j$ for primary fermion fields satisfy the anticommutation relations: $\{\Psi_i,\Psi_j^{\dag}\}=\delta_{ij}$. Then the two point correlators are given as
\bqn
\langle\Psi_i\Psi_j^{\dag}\rangle=C_{ij}, \; \langle\Psi_i^{\dag}\Psi_j\rangle=\delta_{ij}-C_{ij}, \; \langle\Psi_i\Psi_j\rangle=\langle\Psi_i^{\dag}\Psi_j^{\dag}\rangle=0
\eqn
The reduced density matrix of the fermion system can be written in the exponential form \cite{Casini:2009sr}:
\bqn
\rho_V=Ke^{-\mathcal{H}}=Ke^{-\Sigma_VH_{ij}\Psi_i^{\dag}\Psi_j},
\eqn
where $\mathcal{H}$ is the modular Hamiltonian of the system and $K$ is the normalization constant which satisfies $Tr\rho_V=1$. The two point correlators $C_{ij}$ in the region $V$ of space is related to the reduced density matrix $\rho_V$ by the following equation:
\bqn
C_{ij}=Tr(\rho_V\cdot\Psi_i\Psi_j^{\dag}).
\eqn
We can diagonalize the exponent by the Bogoliuvov transformation $d_\ell=U_{\ell m}\Psi_m$, with unitary operator $U$ in order to maintain the anticommutation relation $\{d_i,d_j^{\dag}\}=\delta_{ij}$. We choose $U$ such that $UHU^{\dag}=\{\epsilon_i\}$ is a diagonal matrix and $\epsilon_i$ is the eigenvalue of Hermitian matrix $H$. Using the normalization condition $Tr\rho_V=1$ and the Bogoliuvov transformation, then the reduced density matrix $\rho_V$ can be  rewritten as
\bqn
\rho_V=\prod_\ell\frac{e^{-\epsilon_\ell\cdot d_\ell^{\dag}d_\ell}}{1+e^{-\epsilon_\ell}}.
\eqn
And the relation between $H$ and $C$ can be also rewritten as
\bqn
K\cdot Tr(e^{-\Sigma_VH_{lm}\Psi_l^{\dag}\Psi_m}\cdot\Psi_i\Psi_j^{\dag} )=K\cdot Tr(\prod_\ell\frac{e^{-\epsilon_\ell\cdot d_\ell^{\dag}d_\ell}}{1+e^{-\epsilon_\ell}}\cdot\Psi_i\Psi_j^{\dag} )=C_{ij}.
\eqn
Next we diagonalize the two point correlators $C_{ij}$ by Bogoliuvov transformation, we can obtain
\bqn
\text{diag}\{C_{ij}\}=\prod_{\ell=1}^{N}\frac{1}{1+e^{-\epsilon_\ell}}.
\eqn
We define $C_\ell$ as the eigenvalues of the matrix $\text{diag}\{C_{ij}\}$, then we have
\bqn
\epsilon_\ell=-\log(\frac{1}{C_\ell}-1), \; C_\ell \in (0,1).
\eqn
In terms of the definition of the von Neumann entropy (\ref{1812}), then the entanglement entropy for primary fermion fields of the region $V$ can be written as
\begin{eqnarray}
S_{V}&=&-Tr(\rho_V\ln\rho_V)=-Tr\left(\prod_\ell\frac{e^{-\epsilon_\ell\cdot d_\ell^{\dag}d_\ell}}{1+e^{-\epsilon_\ell}}\cdot\log\left(\prod_\ell\frac{e^{-\epsilon_\ell\cdot d_\ell^{\dag}d_\ell}}{1+e^{-\epsilon_\ell}}\right)\right)\nonumber\\
&=&\sum_{\ell}\left(\log\left(1+e^{-\epsilon_\ell}\right)+\frac{\epsilon_\ell\cdot e^{-\epsilon_\ell}}{1+e^{-\epsilon_\ell}}\right)\nonumber\\
&=&-\sum_{\ell}\left((1-C_\ell)\cdot\log(1-C_\ell)+C_\ell\cdot\log C_\ell\right)\nonumber\\
&=&-Tr\left[\left(1-C_{V}\right)\log\left(1-C_{V}\right)+C_{V}\log C_{V}\right],
\end{eqnarray}
where we have traced two quantum states such as $|0\rangle$ and $|1\rangle$ for primary fermion fields in the second line.

\end{appendix}

\end{document}